\documentclass{PoS}
\newcommand{\vtwo}{v$_2$ }
\newcommand{\vtwonq}{v$_2$/nq }
\newcommand{\qhat}{$\hat\mathrm{q}$ }
\newcommand{\pt}{p$_\mathrm{T}$ }
\newcommand{\ptnq}{p$_\mathrm{T}$/nq }
\newcommand{\mtnq}{m$_\mathrm{T}$/nq }
\newcommand{\etaOverS}{$\eta$/s }
\newcommand{\pions}{$\pi^\pm$ }
\newcommand{\kaons}{K$^\pm$ }
\newcommand{\kzero}{K$^0_\mathrm{s}$ }
\newcommand{\antip}{$\bar\mathrm{p}$ }
\newcommand{\lambdas}{$\Lambda$ }
\newcommand{\phis}{$\phi$ }
\newcommand{\xis}{$\Xi^\pm$ }
\newcommand{\omegas}{$\Omega$ }
\newcommand{\dzero}{D$^0$ }
\newcommand{\dplus}{D$^+$ }
\newcommand{\dstar}{D$^{*+}$ }
\newcommand{\energyLHC}{$\sqrt{\mathrm{s_{NN}}}=2.76$ TeV }
\newcommand{\energyRHIC}{$\sqrt{\mathrm{s_{NN}}}=200$ GeV }
\newcommand{\gevc}{GeV/$c$}

\title{Flow of strange and charm particles in Pb--Pb collisions at \energyLHC measured with ALICE}
\ShortTitle{Flow of strange and charm particles}
\author{\speaker{Carlos Eugenio Perez Lara}\ \ for the ALICE Collaboration\\ %\thanks{A footnote may follow.}\\
  Nikhef\\
  E-mail: \email{cperez@nikhef.nl}}
\abstract{
The ALICE experiment studies Pb-Pb collisions at the LHC in order to investigate the properties of the hot and dense QCD matter at extreme energy densities.
Recent results from ALICE in identified particle flow allow for the exploration of the collective properties of the medium created in heavy-ion collisions.
In this paper, I give special attention to strange and charm particles which probe the medium differently and thus provide new constraints for the study of its properties.
The paper covers results on elliptic flow for \pions, \kaons, \kzero, \antip, \phis, \lambdas, \xis, \omegas, \dplus, \dzero and \dstar measured at midrapidity by ALICE in Pb-Pb collisions at \energyLHC.
I present also the comparison with available models that predict the hydrodynamical evolution of the medium and the energy loss of light and heavy quarks as they travel through.
}
\FullConference{36th International Conference on High Energy Physics,\\
  July 4-11, 2012\\
  Melbourne, Australia}

\begin{document}
\section{Introduction}
A state of deconfined QCD matter is achieved by heavy ion collisions at LHC.
The ALICE experiment has a broad physics program pursuing the systematical study and characterization of such state.
Anisotropic flow\cite{Voloshin:1994} is a well established observable sensitive to the expansion of the thermalized quark matter.
This observable constrains not only the parameters of the equation of state of the produced matter, but also the parameters that describe the energy loss of particles traversing the hot medium.
Furthermore, anisotropic flow provides insight into the mechanisms of hadronization such as coalescence and parton fragmentation
as will be discussed in section 3.

In these proceedings I report the latest measurements of elliptic flow for identified particles in Pb--Pb collisions at \energyLHC measured at midrapidity $|\eta|<0.8$ with the ALICE detector.
In addition a brief description of what we learn from these measurement as well as the comparison to available models is presented.

\section{Constraining the equation of state}
\begin{figure}
  \centering
  \includegraphics[width=19em]{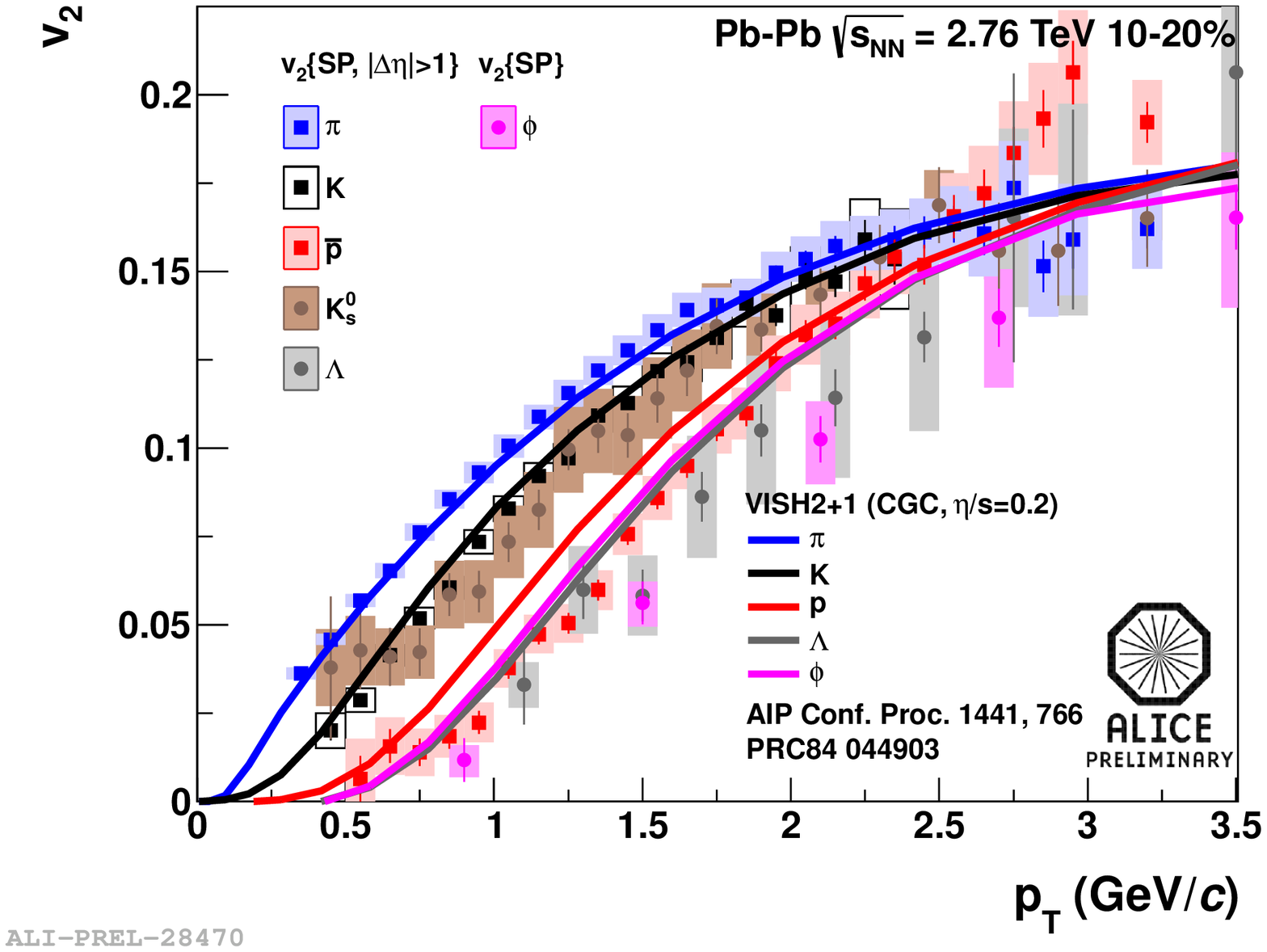}
  \includegraphics[width=19em]{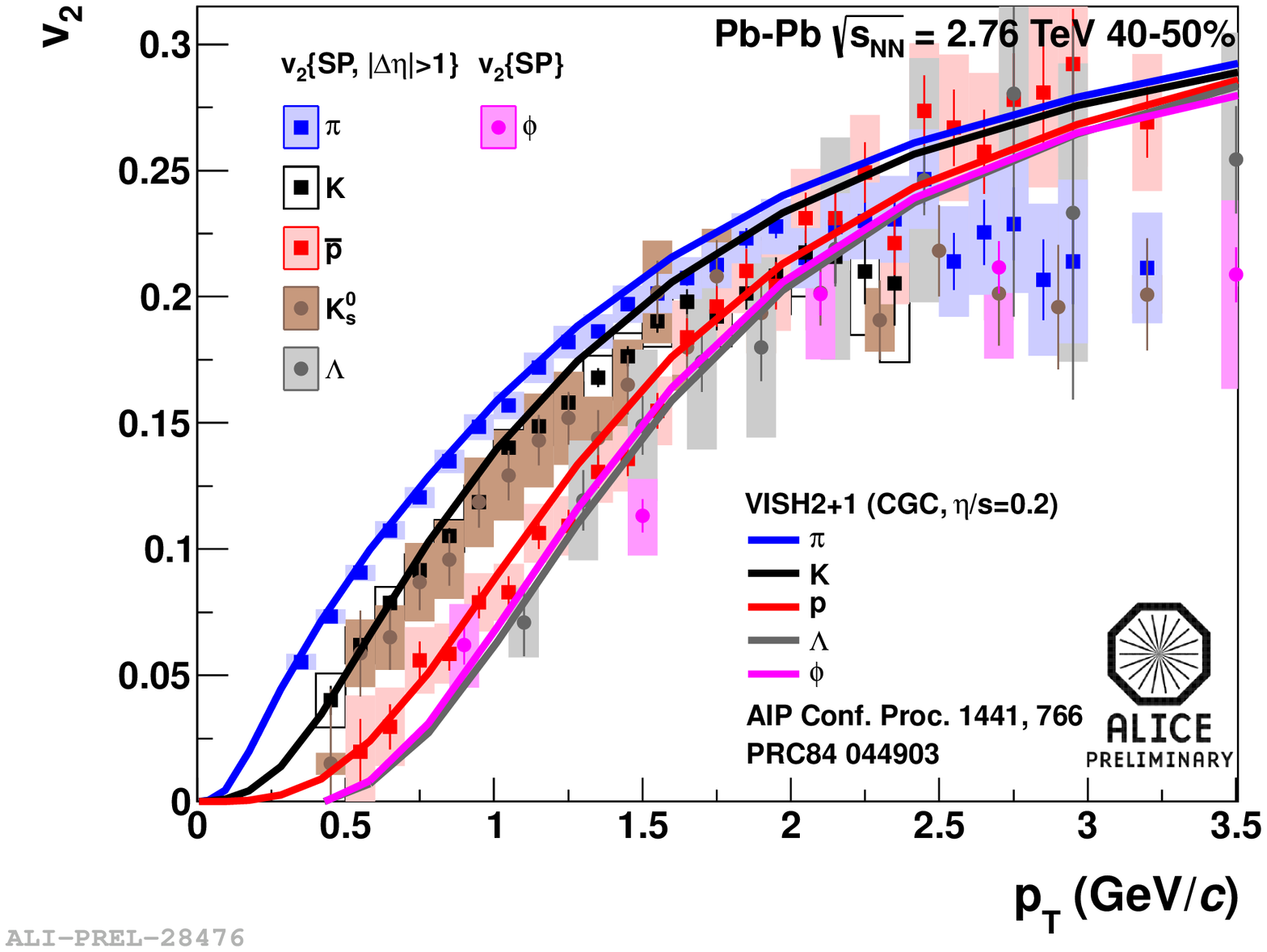}
  \caption{Elliptic flow as a function of transverse momentum for different species. Left figure corresponds to centrality class 10-20\%. Right figure, to centrality class 40-50\% }
  \label{figOne}
\end{figure}

At RHIC energies the evolution of the system from thermalisation to hadronization has been successfully described by hydrodynamics\cite{STAR:2001}\cite{PHENIX:2003}.
At those energies the characteristics of the formed medium resemble those of a liquid with a very low viscosity.
The pressure gradients formed by the anisotropic fireball give rise to an anisotropy in momentum of the produced particles.
The degree of anisotropy in momentum can be experimentally quantified via a Fourier decomposition of its azimuthal distribution.
The elliptic flow \vtwo is the second harmonic and the strongest for almost all collisions.
\vtwo is sensitive to the viscosity scaled by the entropy \etaOverS and to the radial boost that characterise the deconfined state.

There are different techniques to measure elliptic flow.
All of them have different sensitivity to nonflow (azimuthal correlations not originated by the collectivity), multiplicity fluctuations and flow fluctuations.
The identified particle \vtwo measurement presented here was performed using the Scalar Product\cite{ALICE:2010} method, imposing a separation in $\eta$ between the species under study and the particles used to compute the Q-vector\cite{Voloshin:2008}.
The particle identification was made using the Time Projection Chamber and Time Of Flight detectors and the Inner Tracking System.
Charged particles \pions, \kaons and \antip were identified by their specific energy loss and time of flight in the central barrel\cite{ALICE:2004}\cite{ALICE:2006}.
While for decaying particles, the reconstruction was performed based on topology and invariant mass analysis.

Figure \ref{figOne} shows the elliptic flow as a function of \pt for identified particles: \pions, \kaons, \kzero, \antip, \phis and \lambdas
\footnote{The measurement for \xis and \omegas can be found in the conference slides.}.
The results for centrality class\cite{Toia:2011} 10-20\% (40-50\%) is shown on the left (right) together with the viscous hydrodynamical predictions\cite{Shen:2011}.
A clear separation in \vtwo for different species and the expected mass hierarchy is observed at LHC energies.
The lines represent viscous hydrodynamical calculations using MC-KLN initial conditions.
The data can be described using a value of \etaOverS $=0.2$ for the deconfined state.
In spite of the fact that the model reproduces very well the mass hierarchy, it overpredicts \vtwo for the most massive species at very central collisions.
This discrepancy suggest a stronger re-scattering in the hadronic phase for the most central collisions.

It is worth noticing that on both centralities there is a crossing point between mesons and baryons below 3 \gevc.
The results for \xis and \omegas that can be found in the conference slides provide additional support to this conclusion.
The same behaviour was observed at RHIC and has been consider as a signature of quark coalescence\cite{PHENIX:2007}.

\section{Testing the hadronization models}
\begin{figure}
  \centering
  \includegraphics[width=19em]{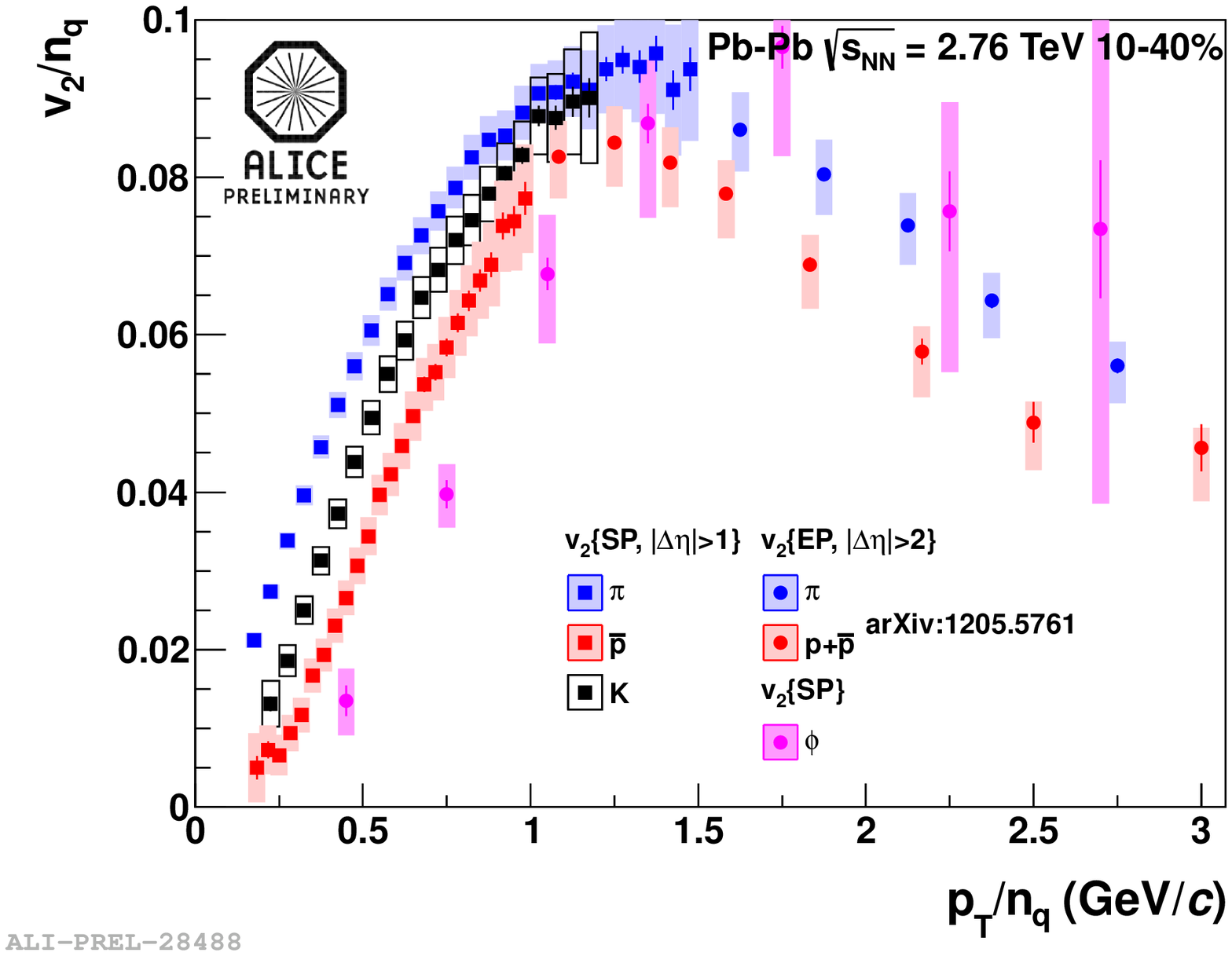}
  \includegraphics[width=19em]{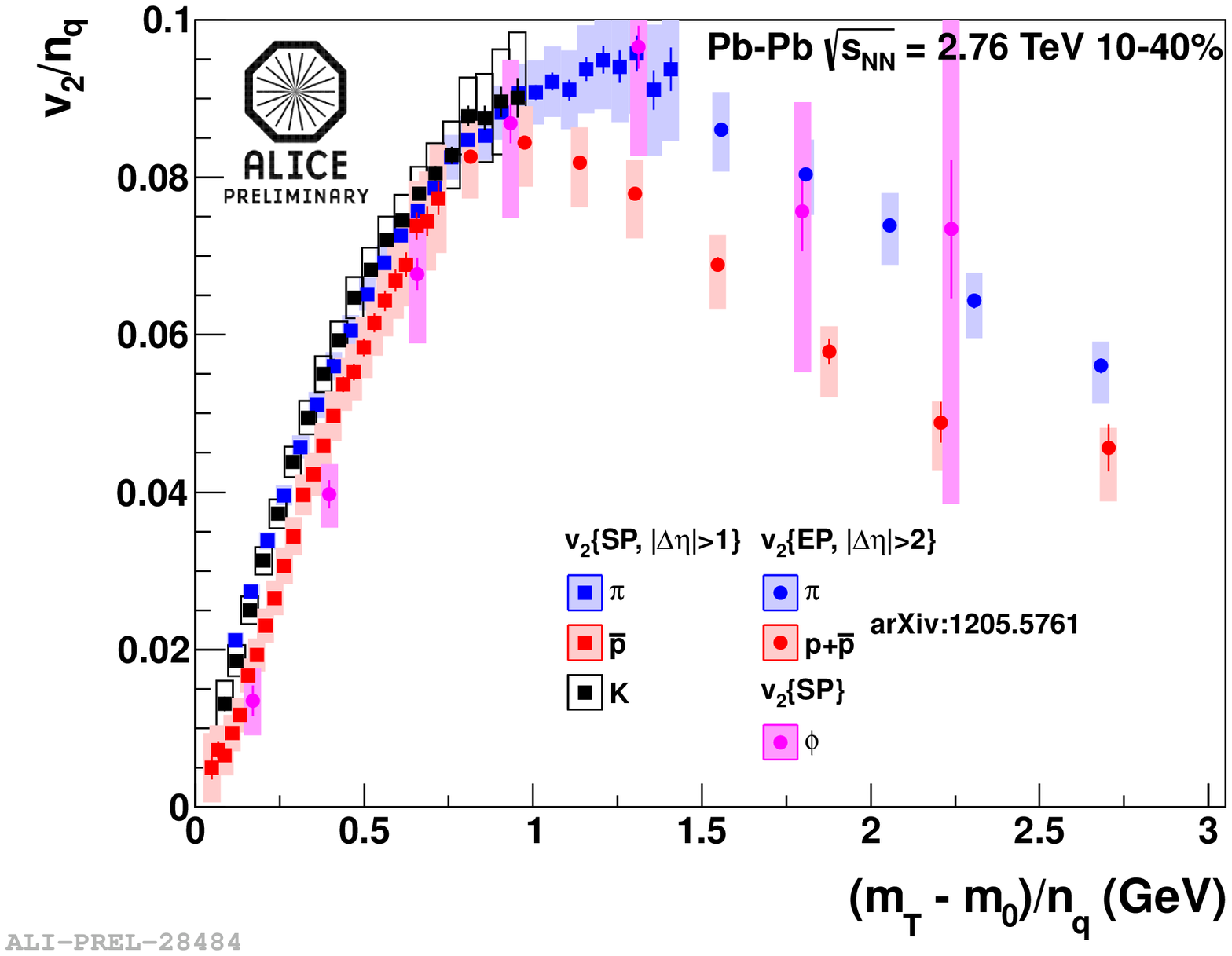}
  \caption{\vtwonq as a function of \ptnq (left figure) and \mtnq (right figure) for different species.}
  \label{figTwo}
\end{figure}

At RHIC energies, the approximate universal scaling\cite{PHENIX:2007}\cite{STAR:2008} of meson and baryon \vtwo -when scaled to their number of valence quaks- provides experimental arguments for a restoration of parton degrees of freedom in the expanding the QCD matter.
A simple quark coalescence production mechanism\cite{Molnar:2003} is sufficient at intermediate \pt -$\sim 3$ \gevc- where there is not major contribution of hydrodynamical boost and parton fragmentation.

Despite the relative success of coalescence at RHIC energies, the elliptic flow for identified particles at LHC shows no universal scaling with the number of constituent quarks.
Figure \ref{figTwo}-left shows the scaled elliptic flow as a function of \ptnq for identified particles: \pions, \kaons, \kzero, \antip, \phis and \lambdas
\footnote{The measurement for \xis and \omegas can be found in the conference slides.}.
Around $~1$ \gevc{} ($~2$ \gevc{} for mesons and $~3$ \gevc{} for baryons), \vtwo for all species does not follow a universal partonic \vtwo as proposed in \cite{Molnar:2003}.

Figure \ref{figTwo}-right shows the scaled elliptic flow as a function of transverse kinetic energy scaled by the number of constituent quarks.
Although the light hadrons cluster together as coalescence predicts, we find a partial disagreement with the coalescence picture due to the accentuated deviation of \antip from light mesons that would point to a breaking of a simple quark coalescence picture.
%different saturation curves.

At higher \pt we expect that the production of jet fragmentation and in-medium energy loss gradually take over.
We found a clear splitting between \pions and \antip at transverse momenta above 3 \gevc.
We believe that these new data can help constrain further the jet fragmentation and in-medium energy loss models.

\section{Probing the medium density}
\begin{figure}
  \centering
  \includegraphics[width=19em]{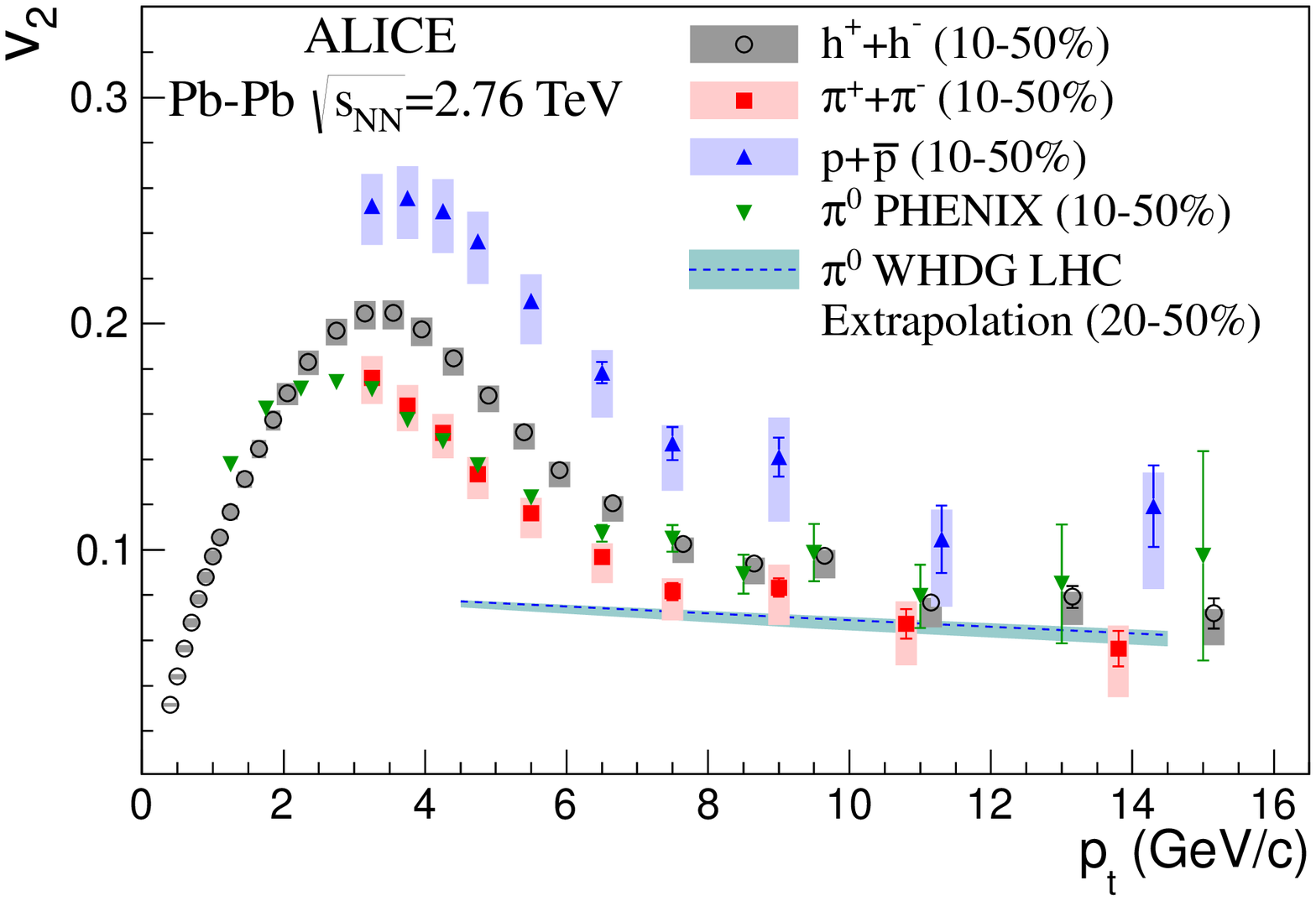}
  \caption{\vtwo as a function of traverse momenta for charged particles, \pions and \antip up to 16 \gevc.
    Figure adapted from \cite{ALICE:2012}.}
  \label{figThree}
\end{figure}

The dynamics of high-pt particles traversing through the medium is currently understood as partons undergoing multiple scattering with scattering centres in the medium and in-medium gluon radiation which depends strongly on the mass of the travelling quark.
The characterisation of such can be computed in terms of diffusion dynamics\cite{Casalderrey:2007}.
The key parameter in such models is the transport coefficient \qhat(t) which encodes information about the expanding dynamical QCD matter.

Important constraints to \qhat come from fits to the measured nuclear modification factor R$_\mathrm{AA}$ for light and heavy particles.
For a geometrically anisotropic expanding medium, high \pt particles are suppressed in medium according to their azimuthal angle.
The measurements of anisotropic flow at high traverse momenta\cite{ALICE:2012} is sensitive to the path length dependent energy loss of particles going through the medium.
Figure \ref{figThree} shows \vtwo for all charged hadrons, \pions and \antip as a function of \pt up to 16 \gevc{} for centrality class 10-50\%.
Above 10 \gevc{} the \vtwo develops a plateau which is small but constant pointing to the relationship between the path-length dependent energy loss for hadrons and the final size and shape of the system.
The energy loss computed by the WHDG model\cite{Horowitz:2012} tuned for LHC energies reproduces the data very well.
This data provides the opportunity to further constrain the current energy loss models.

Moreover, data from $\pi^0$ measured by PHENIX\cite{PHENIX:2010} in Au--Au collisions at \energyRHIC shows a remarkably similar suppression when compare to \pions at LHC energies.

\section{First look at charm mesons}
\begin{figure}
  \centering
  \includegraphics[width=19em]{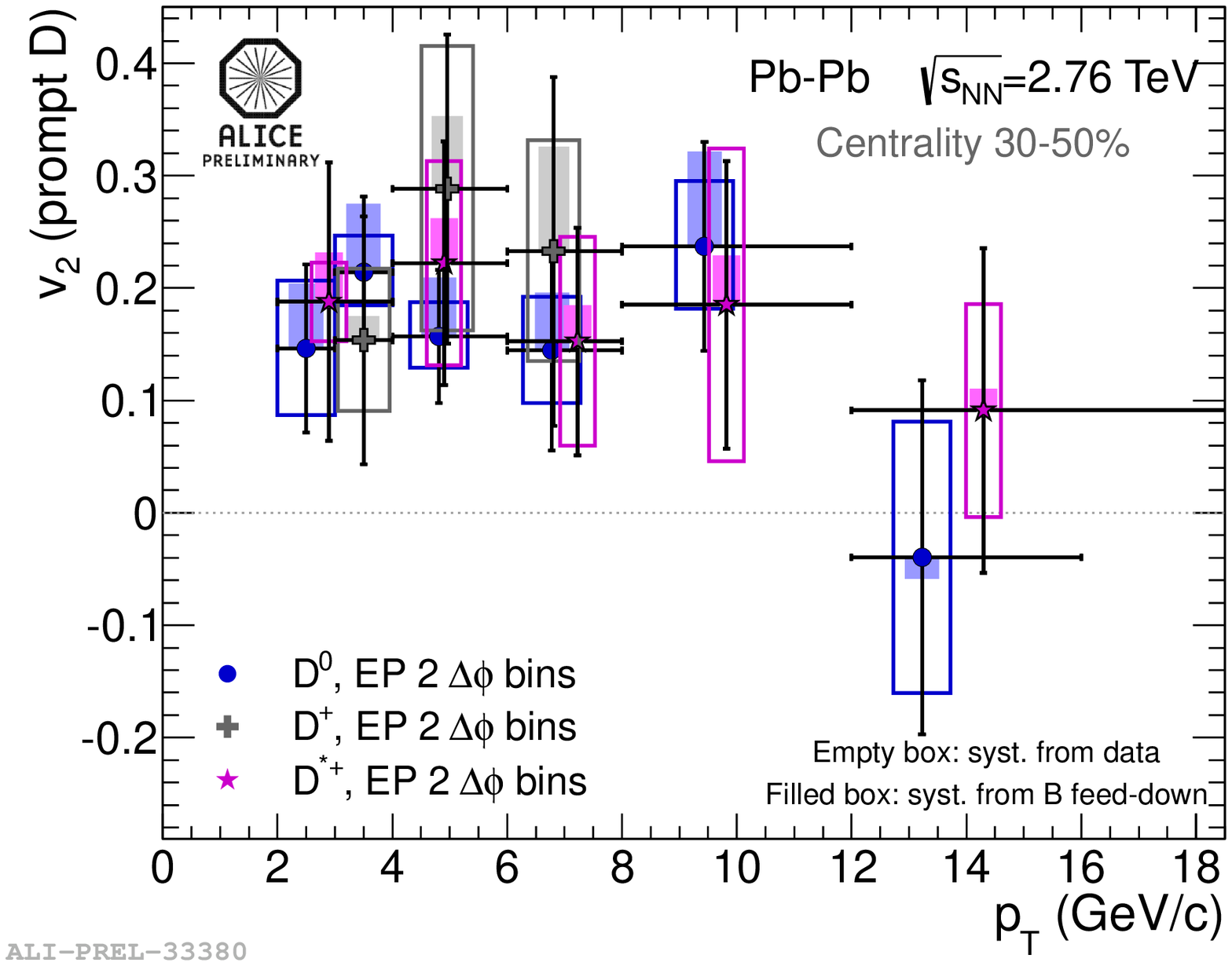}
  \includegraphics[width=19em]{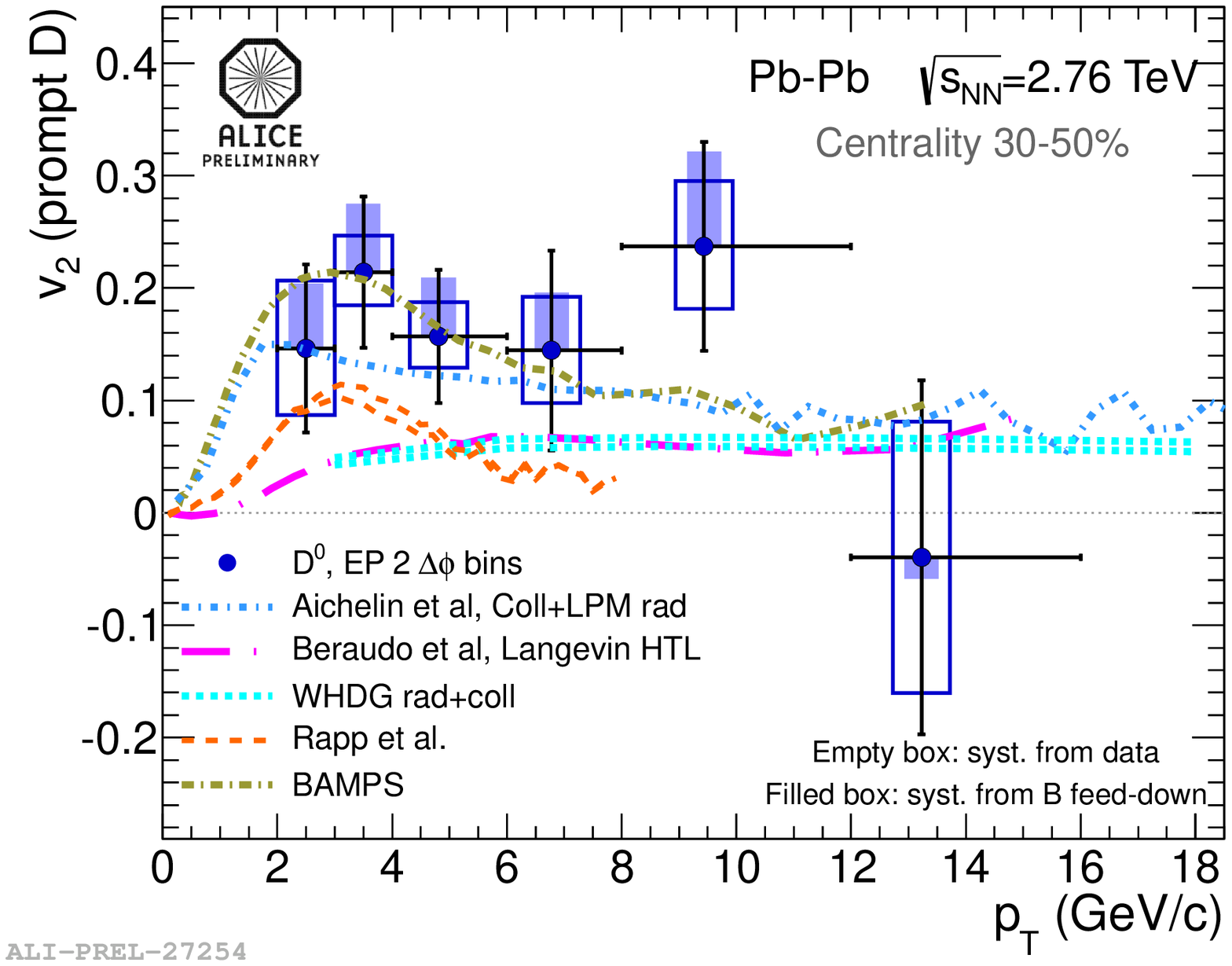}
  \caption{\vtwo as a function of traverse momentum for D mesons (left figure) and comparison to models that describe energy loss in the medium (right figure).}
  \label{figFour}
\end{figure}

Lattice QCD predicts that the deconfined state is reached at temperatures around $\mathrm{T_c} \sim 170$ MeV\cite{LATTICE:2008}.
Temperatures at LHC are predicted to reach $\mathrm{T}_0 \sim 3-5 \mathrm{T_c}$.
At these temperatures the thermal production rate of charm quarks is small.
This is why one can assume that most charm quarks are produced in hard processes at the beginning of the reaction.
That make charm a very insightful probe of the reaction as it is conserved during the whole evolution of the system.

ALICE has made the first open charm \vtwo measurement on fully reconstructed D mesons from their decay into charged hadrons.
The measurement was performed using 9.1 million minimum bias events.
Figure \ref{figFour}-left shows that the measurement of \vtwo for \dzero, \dplus and \dstar give consistent results within errors.
%The \vtwo reported here was measured by the yield asymmetry in- vs out-of-plane, the measurement was validated by two-particle correlation methods as the ones presented in the previous sections.
%Even though different techniques to reconstruct the reported D-meson were used, their differential \vtwo agree well within errors.
The \vtwo reported here was measured by the yield asymmetry in- vs out-of-plane.
The results using two-particle correlation methods\footnote{The results for \dzero \vtwo using scalar product and Q-Cumulants can be found in the conference slides} -as the ones presented in the previous sections- agree well within errors.

In figure \ref{figFour}-right we compare the \dzero \vtwo to five model calculations.
These models have different approaches to the transport of the heavy quarks in the medium and to the hadronic phase.
Some models -like Aichelin\cite{Gossiaux:2012}, Rapp\cite{He:2012} and Beraudo\cite{Alberico:2012}- expand the scattering centres by a hydrodynamic evolution tuned to data.
Models from Aichelin\cite{Gossiaux:2012} and WHDG\cite{Horowitz:2012} have a built-in treatment of in-medium gluon radiation.
The BAMPS\cite{Uphoff:2012} model considers only collisional processes, but mimics the radiative processes by enlarging the cross section to reproduce RHIC R$_\mathrm{AA}$ for heavy flavour electrons.
Aichelin, BAMPS and in less degree Rapp are able to reproduce the measured \vtwo for \dzero at low \pt.
Moreover all models agree in the characterisation of \vtwo for D mesons at high traverse momentum.

\section{Conclusions}

The differential elliptic flow as a function of transverse momentum for \pions, \kaons, \kzero, \antip, \phis, \lambdas, \dzero, \dplus and \dstar was presented up to 16 \gevc.
Light flavour hadron elliptic flow agree with viscous hydrodynamical calculations for \etaOverS $=0.20$ using MC-KLN initial conditions.
Moreover it was found that coalescence is at least partially broken at intermediate traverse momenta.
At high transverse momentum, a small but finite \vtwo was found for \pt higher than 10 \gevc{} which is in agreement with path-length dependent energy loss in medium.
%Finally results for open charm show a high \vtwo down to 2 GeV, while at high traverse momentum the measurement is in agreement with in-medium energy loss calculations.
Finally results for open charm show an indication for \vtwo>0 in the range 2-6 \gevc. Model calculations with charm quark transport and path length dependent energy loss can describe qualitatively the data at low and high \pt, respectively.

\end{document}